\documentclass[prl,byrevtex,balancelastpage,letterpaper,10pt,twocolumn,showpacs]{revtex4}
\usepackage{amssymb}
\usepackage{amsmath}
\usepackage{graphicx}

\newcommand {\be}{\begin{equation}}
\newcommand {\ee}{\end{equation}}
\newcommand {\bea}{\begin{eqnarray}}
\newcommand {\eea}{\end{eqnarray}}

\begin{document}

\title{Doppler cooling a microsphere}

\author{P. F. Barker}
\affiliation{Department of Physics and Astronomy, University College London, WC1E 6BT, United Kingdom}

\begin{abstract}
Doppler cooling the center-of-mass motion of an optically levitated microsphere via the velocity dependent scattering force from narrow whispering gallery mode (WGM) resonances is described.  Light that is red detuned from the WGM resonance can be used to damp the center-of-mass motion in a process analogous to the Doppler cooling of atoms.  Leakage of photons out of the microsphere when the incident field is near resonant with the narrow WGM resonance acts to damp the motion of the sphere.  The scattering force is not limited by saturation, but can be controlled by the incident power.  Cooling times on the order of seconds are calculated for a 20 micron diameter silica microsphere trapped within optical tweezers, with a Doppler temperature limit in the microKelvin regime.

\end{abstract}

 \maketitle \preprint{} \eid{} \startpage{1}
\endpage{}

Doppler cooling has been an extremely successful technique for cooling atomic species to temperatures in the microKelvin regime, opening up new areas in atomic \cite{chu}, molecular \cite{ye}, condensed matter \cite{esslinger}, and many body physics \cite{manybody}.  It has allowed the creation of atomic gases in the quantum regime, including the creation of Bose-Einstein condensates of atomic gases \cite{bose} and Fermi gases \cite{trusty}. More recently there has been considerable interest in the cavity cooling of atoms and molecules because a wider range of particles can, in principle, be cooled as no internal resonance is required \cite{horak}.  A resonance is, however, required in the form of an external optical cavity, and a single atom \cite{rempe}, ion \cite{ion} and atomic ensembles \cite{vuleticens} have been cooled. For molecular and atomic species that cannot be laser cooled, cavity cooling of a trapped species appears attractive because it does not rely on the detailed internal level structure.  Over the last ten years the field of cavity optomechanics has cooled micro- and nano-mechanical objects down to temperatures where the quantum mechanical nature of their motion will soon be apparent  \cite{nanoosc1,nanoosc2, karrai} . Like cavity cooling of atoms and molecules, blue shifted photons are scattered from the cavity with respect to the incident photons, thus extracting energy. In this process, at least one degree of the freedom, such as a cavity mirror, or intracavity membrane, is damped or cooled by interaction with the cavity field \cite{reviewopto, karrai, resolvedsideband,membrane, kippenberg}.  An important system of this type is the cooling of the internal mechanical modes of a high Q, WGM resonator formed by a toroidal or spherical structure \cite{kippenberg1}.  Very recently there have been proposals to cool optically levitated particles using cavity cooling \cite{barker,chang,cirac}. This is attractive because the ability to levitate the particle isolates it well from the environment, increasing the prospects of cooling the center-of-mass motion to its quantum ground state.  While this latter scheme is attractive for cooling nanoparticles, it does not appear to be practical for larger particles, which would significantly perturb the cavity field reducing and potentially inhibiting cooling.  

In this letter we describe a hybrid scheme that links laser Doppler cooling and cavity cooling.  This scheme differs from cavity cooling where the particle is cooled within the cavity, or in optomechanics, where part of the cavity is cooled. Instead, we cool the whole cavity in a process analogous to Doppler cooling where the required frequency dependent scattering force is provided by the high Q, WGM of the microsphere. 

\begin{figure}[h]
\vspace{-10pt}
\begin{center}
\includegraphics[scale=0.3,angle=90]{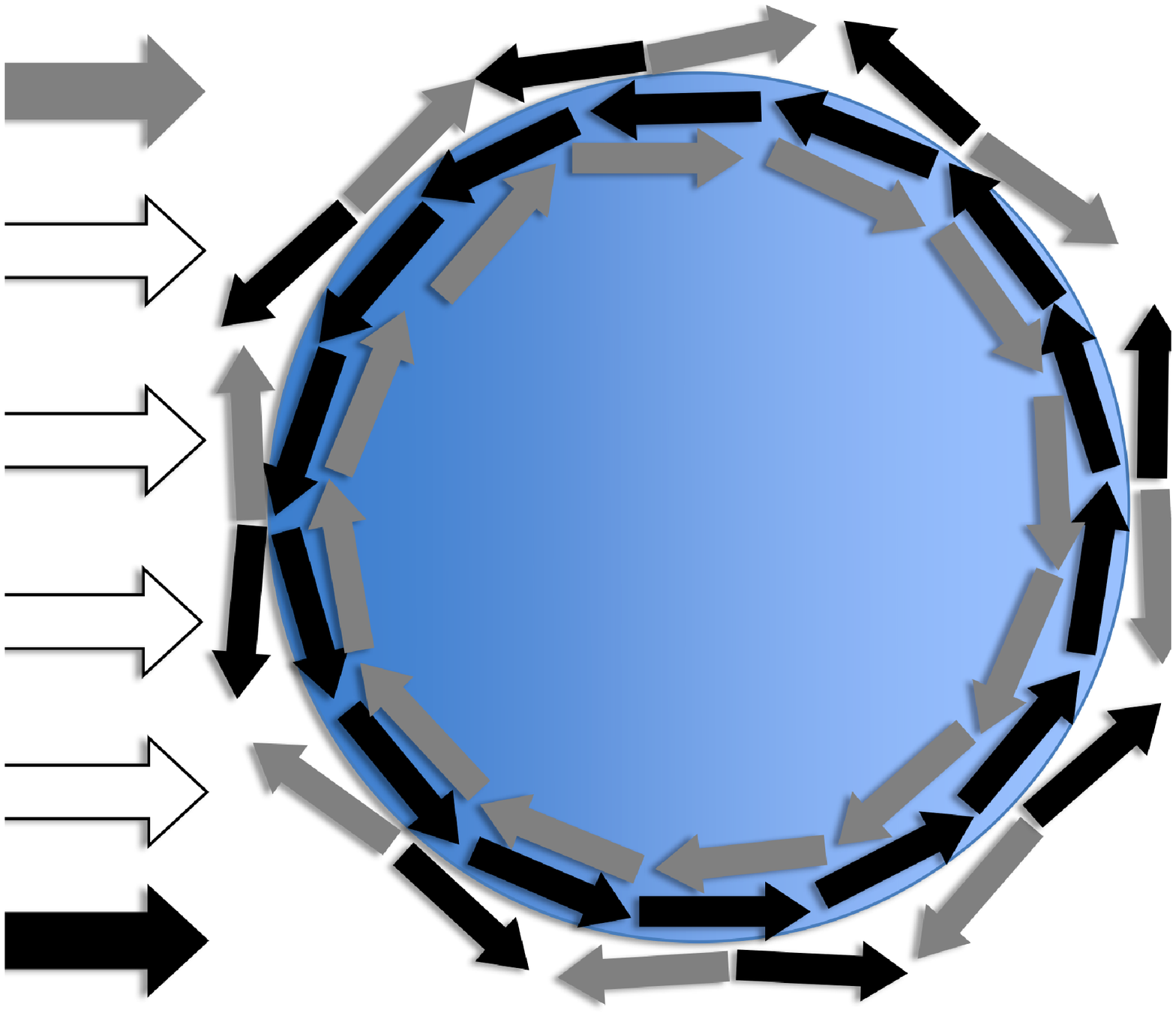}
\caption{(Color online)  The propagation of counter propagating light rays (black and grey) from an incident plane wave within a microsphere on a whispering gallery mode resonance. The white rays are those from the incident field which impinge on the sphere but do not couple into the WGM.  The corresponding isotropic leakage of photons from the sphere is analogous to the average isotropic spontaneous emission of photons from an atom following excitation. } \label{mirror_ring}
\end{center}
\vspace{-10pt}
\end{figure}

Whispering gallery modes or morphology dependent resonances occur in cylindrical and spherical dielectric particles which act as high Q ($>$10$^8$) optical cavities for light that is coupled into the sphere and that propagates by total internal reflection around its annulus. Figure 1 illustrates, from a geometrical optics perspective, the propagation of trapped rays of light from a incident plane wave within one plane through the microsphere. In spherical particles, the excitation of these WGM resonances increases the scattering cross-section and therefore the radiation pressure that can be applied to them. The highest Q's can be achieved when the radius of the microsphere far exceeds the wavelength of the light coupled into it. The spectral widths of these resonances can vary considerably depending on wavelength and the size of the sphere but, like atomic resonances, they can have spectral widths of a few MHz, which are typically limited by weak absorption or by Rayleigh scattering within the sphere.  For these modes the sphere can be seen as a high Q spherical ring-like cavity. The effect of these resonances on the radiation pressure forces was first observed in the early work on levitating spheres by Ashkin and Dziedzic \cite{ashkinresonances}.  Although the coupling of light into these modes is not efficient for a free space propagating optical field, it was observed that the scattering cross-section, and also the optical force from levitation experiments, is enhanced when the incident light is resonant with these modes. Although not discussed by them in this context, the very narrow resonances could be used to damp the motion of the spheres via the Doppler effect, which transforms a frequency dependent force to a velocity dependent force. It is stressed that it is the damping of the center-of-mass motion of the whole microsphere resonator structure and not the internal degrees of freedom in spherical and toroidal resonators which has previously been considered.

To illustrate the forces on a resonator we first consider light incident on an idealized Fabry-Perot resonator, as shown in figure 2a, which is moving towards the beam. We consider the simple case where there are no losses in the mirrors or the medium between them and the reflectivity of each mirror is close to unity. The force on the cavity in a vacuum in the direction of propagation of an incident plane wave is calculated from the incident ($P_\mathrm{i}$), reflected ($P_\mathrm{r}$) and transmitted power ($P_\mathrm{t}$), and is given by $F =1/c(P_\mathrm{i}-P_\mathrm{r}-P_\mathrm{t})$. When the incident light is not near the cavity resonance almost all the incident light is reflected from the cavity and the force is at its maximum $F\approx2P_\mathrm{i}/c$. On resonance all light is transmitted and $F\approx0$. The force is frequency dependent near the cavity resonance and is shown in figure 2a.  The resonance is inverted when compared to an atomic resonance, such that through the Doppler effect more force would be felt by the resonator when it is moving towards an incident field that is blue detuned with respect to the resonance.  The WGM's in a microsphere are, however, more like the ring cavity as shown in single plane in figure 1. Two rays near the annulus of the sphere are coupled and counter propagate around the sphere. To understand how this type of resonator is affected by radiation pressure we consider a more simplistic ring cavity of eight mirrors shown in figure 2b. Two rays of equal power from an incident field are coupled into the side mirrors.  Any other rays from an optical field that are not resonant because of their angle of incidence or position on the cavity, will be reflected and/or refracted producing a force that will not be strongly frequency dependent.  We again consider the incident, reflected and transmitted light at each mirror, and the resulting forces due to the change in momentum, assuming that all mirrors have the same reflectivity and transmissivity with no losses. Unlike the Fabry-Perot resonator the transmitted rays are distributed evenly in a plane. By symmetry, the resulting forces around the ring due to the transmitted rays cancel each other out on average, just as in the case of isotropic spontaneous emission from an excited atom. The force in the $y$-direction is then only due to the reflected and the incident field. Off resonance all the incident light is reflected and there is no net force in the $y$-direction.  The force only acts to maintain the sphere position along the $x$-axs.  On resonance all light is transmitted and the net forces act to push the sphere in $y$ direction, while maintaining the sphere on in its position on the $x$-axis.  Figure 2b also shows a plot of the force in the $y$-direction due to these two rays derived from the reflection of an $n$ mirrored cavity as function of frequency, where $F=2/c(\vec{P_\mathrm{i}}-\vec{P_\mathrm{r}})\cdot  \hat{y}$ and $P_\mathrm{r}=\frac{P_i}{c} |(r-\frac{(r-r^3)e^{-ikL}}{1-r^ne^{-ikL}})|^2$ and $r$ is the amplitude reflection coefficient \cite{seigman}.  In contrast to the Fabry-Perot cavity the force due to radiation pressure is maximised on resonance, just as in the atomic case, and the sphere acts in this respect like a large two level atom.  
\begin{figure}[h]
\hspace{-40pt}
\begin{center}
\includegraphics[scale=0.32,angle=90]{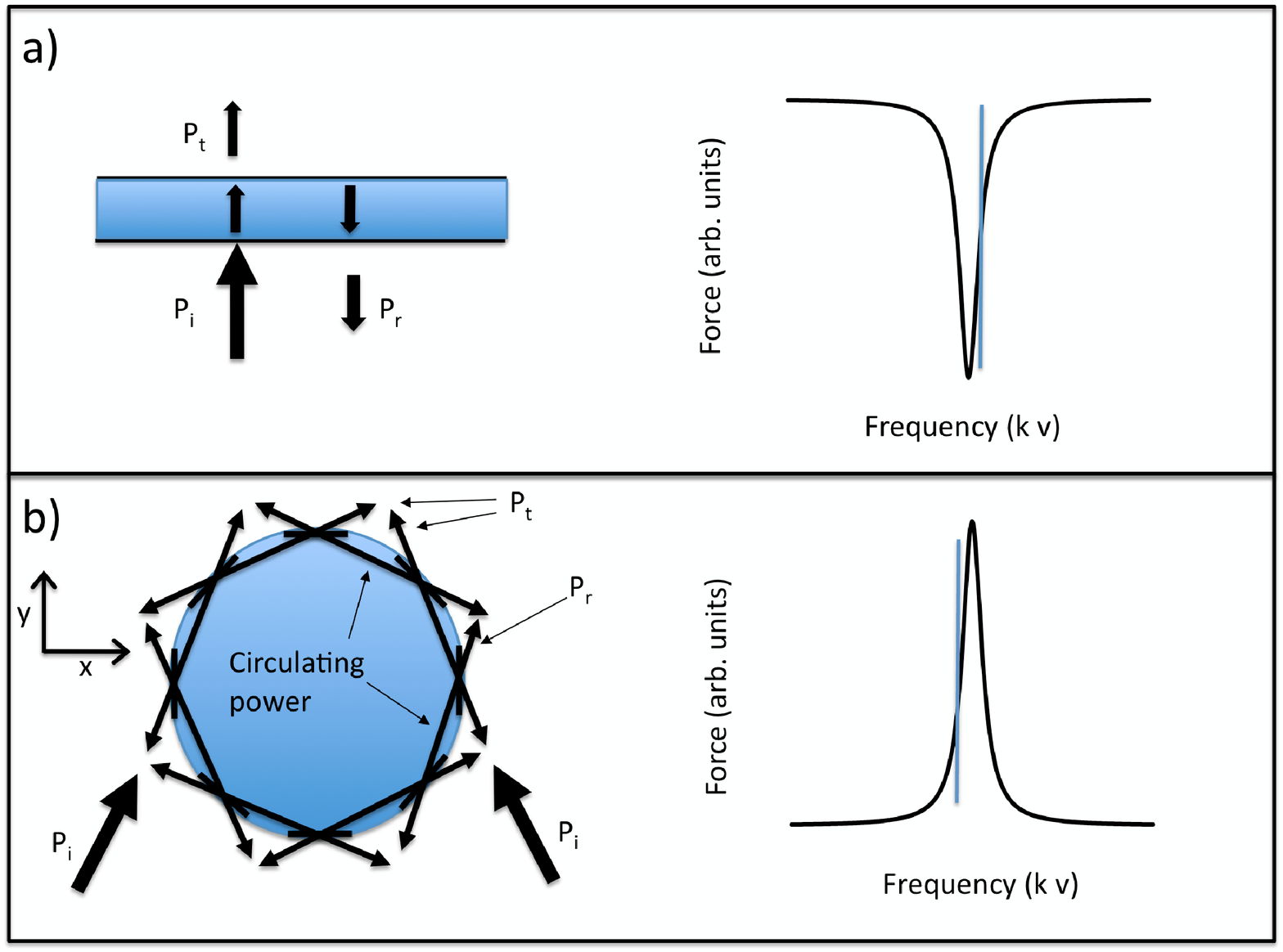}
\caption{(Color online) a) A diagram illustrating the incident ($P_\mathrm{i}$), reflected ($P_\mathrm{r}$) and transmitted ($P_\mathrm{t}$) power from a Fabry-Perot cavity in a vacuum. Also shown is a diagram illustrating the variation in force due to radiation pressure as a function of frequency near the transmission peak. The vertical line represents the detuning with respect to resonance, which would preferentially damp motion of the cavity towards the incident field.  b) A diagram of an eight mirrored ring cavity to illustrate the effect of radiation pressure force on WGM-like ring cavity which approximates the whispering gallery modes of a microsphere.  Also shown is a diagram illustrating the variation in force in the $y$-direction as a function of frequency near resonance with the vertical line representing the red detuning required to damp the motion of the resonator moving in the -ve $y$-direction. The Lorentzian force profile, which peaks on resonance, is analogous to an atomic resonance in which the scattering force is peaked on resonance. } \label{wgm}
\end{center}
\vspace{-20pt}
\end{figure}        

The discussion above served to illustrate the basic physics behind the radiation pressure at resonance. We now calculate a more accurate force due to radiation pressure from an incident plane wave and the scattered light fields via Lorenz-Mie theory. For plane wave illumination the radiation pressure on a non-absorbing sphere in vacuum, with size parameter $x = \frac{2 \pi}{ \lambda}a $, where $a$ is the sphere radius and $\lambda$ is the wavelength of light, is given by $F=\frac{P}{c} Q_\mathrm{rad}$ where $P$ is the beam power and $c$ is speed of light. The normalised radiation pressure cross-section is given by

 $Q_\mathrm{rad}=Q_\mathrm{ext}-4/x^2\displaystyle \sum_{n=0}^{\infty} \lbrace \frac{n(n+2)}{n+1}\mathrm{Re}(a_n a_{n+1}^*+b_n b_{n+1}^*)+\frac{2n+1}{n(n+1)}\mathrm{Re}(a_n b_n^*)\rbrace$
\\where $Q_\mathrm{ext}=2/x^2\displaystyle \sum_{n=0}^{\infty}  (2n+1) Re(a_n+b_n)$. 
The values for the Mie coefficients $a_n$ and $b_n$ can be found from standard texts on Mie scattering \cite{bohren}, where $n$ represents the nth partial wave for the $a_n$ and $b_n$ modes respectively. The Mie resonances, or WGM's, for each can be found by solving for Im$(a_n)=0$ and Im$(b_n)=0$. The mode order $l$ are the roots of the partial wave of mode number $n$, with the lowest order $l=1$ producing the narrowest resonance for the each mode number. Figure 3 is a plot of the radiation pressure force calculated for an incident power of 10 mW and size parameters from 39  to 41. A similar plot is obtained for a nearly collimated Gaussian beam using generalised Lorenz-Mie theory\cite{ott}.  A range of resonances in the scattering force can be observed with very different widths. The narrowest are not well resolved in this low resolution calculation and are indicated by the solid vertical lines. Note that the size parameter can represent a variation in laser wavelength/frequency or a variation in particle radius $a$. The inset graph is a calculation at higher resolution on an expanded scale for the $a_n$( $n=52$ and $l=1$) resonance for size parameter $x$ = 40.62425. This would correspond to a 10 micron radius sphere illuminated with light at approximately 773 nm. The scale has been converted to frequency with respect to the line centre of the resonance, which has a Lorentzian profile with a half width $\delta= 2\pi \times$32 MHz. On resonance the force is 18.5 pN, which on this narrow frequency scale is offset by a constant force of 14.3 pN due to loight that is not reosnant with the WGM's. The force near any of these narrow Mie resonances can be approximated by $F_\mathrm{rad}=\frac{P_0}{c}+\frac{P_\mathrm{p}}{c}\frac{ \delta^2}{(\omega-\omega_0)^2+\delta^2}$, where $\frac{P_0}{c}$ is the constant over the frequency range considered and $\frac{P_p}{c}$ is the peak resonant force.  Here $\delta$ is the half-width-half-maximum line width and $\omega$ is the frequency of the light that is detuned from the resonant frequency $\omega_0$. The force is dependent on the velocity, $v$, of the microsphere via the Doppler effect and is given by $F_{rad}=\frac{P_0}{c}+\frac{P_p}{c} \frac{ \delta^2}{(\Delta \pm k v)^2+\delta^2} $ where $\Delta$ is the detuning from resonance of the stationary sphere. For the small velocities expected of a trapped microsphere, the force can be expanded about $v =0$ to give $F_\mathrm{rad}=\frac{P_\mathrm{0}}{c}+\frac{P_p}{c} \frac{ \delta^2}{\Delta^2+\delta^2}+\frac{P_\mathrm{0}}{c} \frac{\mp 2 k \Delta \delta^2}{(\Delta^2+\delta^2)^2} v$.  Unlike laser Doppler cooling of atoms there is no saturation of the cooling force.  Therefore, in principle, the damping rate can be determined by the power of the light field $P_\mathrm{p}$, although in practice additional heating, introduced by absorption, is likely to shift the resonance.  

\begin{figure}[h]
\includegraphics[scale=0.35]{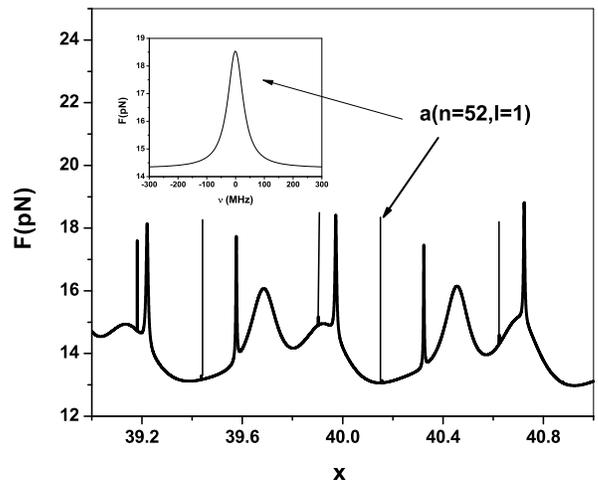}
\caption{(Color online) Calculated force on microsphere for size parameters from $x$ = 39 - 41. This corresponds to a sphere with a radius of 10 $\mu$m illuminated by a plane wave of approximately 773 nm. Inset is a higher resolution of plot of the force as a function of frequency for the $a_n$( $n=52$ and $l=1$) resonance corresponding to a size parameter $x$ = 40.62425.} \label{forcehighandlow}
\end{figure}
We initially consider a single sphere cooled by a 1-D optical molasses using two counter-propagating fields. Here the force on the particle is $F=\beta v$, where $\beta=\frac{4 k P_\mathrm{p} \Delta \delta^2}{c(\Delta^2+\delta^2)^2} $. The e$^{-1}$ velocity damping time or cooling time is $\tau \approx (\beta /m)^{-1}$.  Using $P_0$=100 mW, beam and mass $m=4\times10^{-12}$ kg corresponding to a =10 $\mu$m SiO$_2$ sphere, gives a characteristic cooling time of $\tau=5.7$ s for $\delta$ = 2$\pi\times$32 MHz HWHM resonance or 178 ms for a 2$\pi\times$1 MHz resonance width.  Like laser Doppler cooling of atoms, this will lead to a 1-D cooling limit based on a balance between the average energy damping rate or cooling power expressed as $\langle P \rangle =\beta \langle v^2 \rangle$ and heating by diffusion with the diffusion constant given by $D=\langle \dot{p}^2 \rangle =\hbar^2 k^2 \Gamma_{sc}$ where $\Gamma_{sc}$ is the frequency dependent scattering rate given by $\Gamma_\mathrm{sc}=\frac{P_\mathrm{p}}{\hbar \omega} \frac{ \delta^2}{(\Delta+ k v)^2+\delta^2}$. In 1-D where $1/2k_bT$=$1/2 m\langle v^2 \rangle$ and where $\Delta=\delta$, the Doppler cooling limit is the same as for atomic systems and is given by $k_bT\approx\hbar\delta$.  For the 32 MHz resonance this corresponds to a temperature of 760 $\mu$K, and 24 $\mu$K for the 1 MHz resonance. 

An efficient optical trap for dielectric microspheres is produced by two diverging counter-propagating fields in a two-beam fiber trap \cite{mcgloin}.  A single collimated beam that is red detuned from resonance could be used to cool the trapped sphere in analogy to the Doppler cooling of a trapped ion.  As microspheres are often trapped in air we compare the damping of their motion due to gas viscosity and to Doppler cooling. For an optically sphere trapped within a viscous medium such as air, the equation of motion along one axis of the trap is given by
\begin{equation}
mx''(t)=-\omega_0^2 x(t) +(\beta_\mathrm{t}-\Gamma_0) x'(t)+ F_0+F_\mathrm{f}(t),
\end{equation} 
where $\omega_0= \sqrt{\kappa/m}$ is the trap frequency, $\kappa$ is the spring constant of the trap. The drag coefficient for damping of a spherical particle by gas viscosity ($\eta$) is $\Gamma_0=6\pi\eta a$ and by Doppler cooling is $\beta_\mathrm{t}=\frac{2 k \Delta \delta^2}{c(\Delta^2+\delta^2)^2} $. The microsphere will also be subjected to a frequency independent offset force $F_0=\frac{P_0}{c}+\frac{P_\mathrm{p}}{c} \frac{ \delta^2}{\Delta^2+\delta^2}$ and a time-varying Langevin force $F_f(t)$.  At 288 K, the damping in air is given by $\Gamma=3.4\times10^{-9}$ kgs$^{-1}$ while the maximum value of the optical damping coefficient for a 100 mW incident beam is only three orders of magnitude less at $\beta_t=6.7\times10^{-12}$ kgs$^{-1}$. These values are approximately equal at a pressure of 15 mTorr, based on the drag coefficient of a sphere in the free molecular flow regime where $\Gamma_0=(4/3+3\pi/16)\pi\rho \langle v \rangle a^2$,  $\langle v \rangle$ is the mean velocity of the gas particles and $\rho$ is the gas density \cite{epstein}.  The optical damping is six orders of magnitude greater than that due to the background gas at 1$\times10^{-6}$ torr.  

A typical spring constant of an optical trap is $k$ = 5$\times$10$^{-5}$ and $\omega_0$=2 $\pi\times$ 740 Hz and, using the mass calculated from a spherical silica sphere of radius $a$ = 10 microns, the motion should be damped on the order of seconds. This does not correspond to sideband resolved cooling, but this could be accomplished by cooling on a resonance that is narrower than the trap frequency or by increasing the trap frequency via an increase in light intensity used to trap the particle. The equation of motion above also applies to a microsphere attached to a cantilever and Doppler cooling of the microsphere can be used to cool the cantilever motion. Here the spring constant of the cantilever is much larger than an optical trap ($< $80 N/m) with oscillation frequencies below 1 MHz typically in the 100 kHz range. A cantilever of $ k = $ 77 N/m and resonant frequency of 1 MHz, has an effective mass of $2\times10^{-12}$ kg which is similar to the mass of the 10 $\mu$m sphere considered above. Sideband resolved cooling therefore also appears feasible in this optomechanical system.  

As the resonance frequency for any microsphere is critically dependent on its radius, the resonance condition for each sphere must be found in order to begin cooling it. This requires that the laser must be offset locked to the resonance. Locking a laser to WGM resonance has been demonstrated \cite{locking} and this will be more easily found for the microsphere cantilever system since the sphere is permanently attached to the cantilever. Silica microspheres have been trapped in vacuum for up to half an hour in vacuum at 10$^{-6}$ torr with the time limited by radiometric heating from absorption of the laser light at 514.5 nm \cite{ashkin_lev}. This time could be considerably improved by using low loss SiO$_2$ spheres, as well as using trapping and cooling light in the low-loss wavelength window around 1-1.5 microns where narrow bandwidth low noise lasers are available.  Finding and subsequently locking the laser to the appropriate WGM could be carried out while the particle is trapped in air, which we have demonstrated can be trapped in excess of 5 hours. Once locked to the red side of the resonance, the air could be pumped out and the microsphere cooled to its Doppler limit.   
A microsphere has internal mechanical resonances which can be excited by radiation pressure acting on the internal surfaces of the sphere when light is coupled in the WGM\cite{microsphere_vib}. This will induce motional sidebands on the microspheres WGM resonance. When light is red detuned from resonance it has been shown that this motion can also be cooled, and thus it may be feasible to cool both the internal and external degrees of freedom the microsphere. This may help to ameliorate the very small residual absorption of light which will act to heat the sphere.  

A method for cooling the center-of-mass motion of a microsphere using the velocity dependent force inherent in the whispering gallery modes was presented. This type of Doppler cooling has much in common with laser cooling of trapped atoms and ions.  A 3-D optical molasses could be used to cool a microsphere in all three dimension or by a single beam when the sphere is trapped using optical or electrostatic fields or attached to a cantilever. Such a scheme may also be used to sympathetically cool optically bound, co-trapped particles which do not possess whispering gallery mode resonances.  Like laser cooling ultimate temperatures in the microkelvin range appear feasible with cooling times on the order of seconds. 


\end{document}